%% file: ms.tex
\documentclass[twocolumn,letterpaper,8pt]{article}
\pdfoutput=1

\input{preamble}

\author{
   {\textrm J. Aaron Pendergrass} \\
 Johns Hopkins University\\
 Applied Physics Laboratory\\
 \texttt{aaron.pendergrass@jhuapl.edu}
 \and
 {\textrm Sarah Helble} \\
 Johns Hopkins University\\
 Applied Physics Laboratory\\
 \texttt{sarah.helble@jhuapl.edu}
 \and
 {\textrm John Clemens} \\
 Johns Hopkins University\\
 Applied Physics Laboratory\\
 \texttt{john.clemens@jhuapl.edu}
 \and
 {\textrm Peter Loscocco} \\
 National Security Agency\\
 \texttt{loscocco@tycho.nsa.gov}
}

\title{Maat: A Platform Service for Measurement and Attestation}
\date{}

\begin{document}
\maketitle
\begin{abstract}
     Software integrity measurement and attestation (M\&A) are
     critical technologies for evaluating the trustworthiness of
     software platforms. To best support these technologies, next
     generation systems must provide a centralized service
     for securely selecting, collecting, and evaluating integrity
     measurements.  Centralization of M\&A avoids duplication,
     minimizes security risks to the system, and ensures correct
     administration of integrity policies and systems.  This paper
     details the desirable features and properties of such a system,
     and introduces \sysname, a prototype implementation of an M\&A
     service that meets these properties. \sysname\ is a platform
     service that provides a centralized policy-driven framework for
     determining which measurement tools and protocols to use to meet
     the needs of a given integrity evaluation. \sysname\ simplifies
     the task of integrating integrity measurements into a range of
     larger trust decisions such as authentication, network access
     control, or delegated computations.
\end{abstract}

\input{introduction}
\input{requirements}
\input{related-work}
\input{architecture}
\input{deployment}
\input{future-work}
\input{conclusion}

\bibliography{bibliography}
\bibliographystyle{ieeetr}

\end{document}

%% file: preamble.tex
\usepackage{color}
\usepackage{graphicx}
\usepackage{subcaption}
\usepackage{url}
\usepackage{makecell} 
\usepackage{microtype}
\usepackage{booktabs}
\newcommand\anote[1]{{\color{red}[JAP: #1]}}
\newcommand\snote[1]{{\color{magenta}[SCH: #1]}}
\newcommand\jnote[1]{{\color{blue}[JMC: #1]}}
\newcommand\slnote[1]{{\color{blue}[SPL: #1]}}

\renewcommand\anote[1]{}
\renewcommand\snote[1]{}
\renewcommand\jnote[1]{}
\renewcommand\slnote[1]{}

\newcommand\sysname[0]{Maat}

\renewcommand{\rothead}[2][90]{\makebox[1ex][c]{\rotatebox{#1}{\makecell[c]{#2}}}}%


%% file: introduction.tex
\section{Introduction}

System integrity is an increasingly important and often overlooked
input to security decisions.  Integrity measurements (or 
\textit{evidence}) can be collected to verify that (1) the correct software
or hardware platform is being used and (2) the system is in a valid
state.  The increasing use of a diverse set of integrity information
to inform security decisions motivates the development of a
centralized, comprehensive, and flexible system-level service for
selecting and producing these types of measurements.  The primary
contributions of this paper are (1) a design for an integrity
measurement and attestation service to evaluate the integrity of local
or remote software components; (2) a description of our prototype
implementation of this design: \sysname, named after the Egyptian
goddess of truth, balance, and justice; and (3) an illustrative
discussion of how \sysname\ can be applied to integrate M\&A with
platform management systems and network-scale middleware.

Trust decisions are common in today's computing environments.  For
example, when logging into an online banking site, users must trust
the integrity of both their local software and the software running
the bank's website. When outsourcing computation to a cloud computing
provider, users must trust that the cloud infrastructure will
faithfully execute their software without allowing third parties to
interfere with or observe their actions. When a client joins a
network, a mutual integrity decision is made: the user generally must
trust the integrity of the services provided by the network, such as
DNS configuration, and the network operator generally trusts the
client and grants access to internal services, such as a local file
sharing server that would not be addressable from the global internet.
No single set of integrity evidence can be used to justify trust in
all of these scenarios. In each case, it is necessary to balance
the trusting party's desire for a complete evaluation, and the trusted
party's desire to limit disclosure of sensitive data.

\sysname\ provides a central framework
and standard application programming interface (API) for policy-driven selection
of measurement utilities and
attestation protocols suitable for a wide variety of 
platform trust decisions.
This centralization is critical to ensure correct administration of
integrity policies and systems, rather than independently managing a 
multitude of disparate integrity measurement systems.

The next subsection gives a general overview of the goals of
M\&A. Section \ref{sec:properties} describes specific desirable
properties for M\&A. Section \ref{sec:related-work} examines related
work against these properties. Section \ref{sec:architecture}
describes the \sysname\ framework. Section \ref{sec:deployment}
describes three example use cases for \sysname\ including experience
in a production environment. Sections \ref{sec:future-work} and
\ref{sec:conclusion} conclude by describing areas of future work and
summarizing this paper's contributions.

\subsection{M\&A Background}

Integrity measurement systems (IMS) provide mechanisms for determining
what software is installed or running on a platform, and validating
the on-disk system configuration and current state of running
software. These systems necessarily include a \emph{measurement agent}
that collects \emph{evidence} describing the state of a \emph{target},
and an \emph{appraiser} that evaluates the integrity of the target
based on the evidence produced. The \emph{attestation protocol}
defines how the evidence from the target is bundled, transferred, and
presented to the appraiser.
\begin{figure*}[t]
     \begin{center}
     \begin{subfigure}{0.22\textwidth}
          \includegraphics[width=\textwidth]{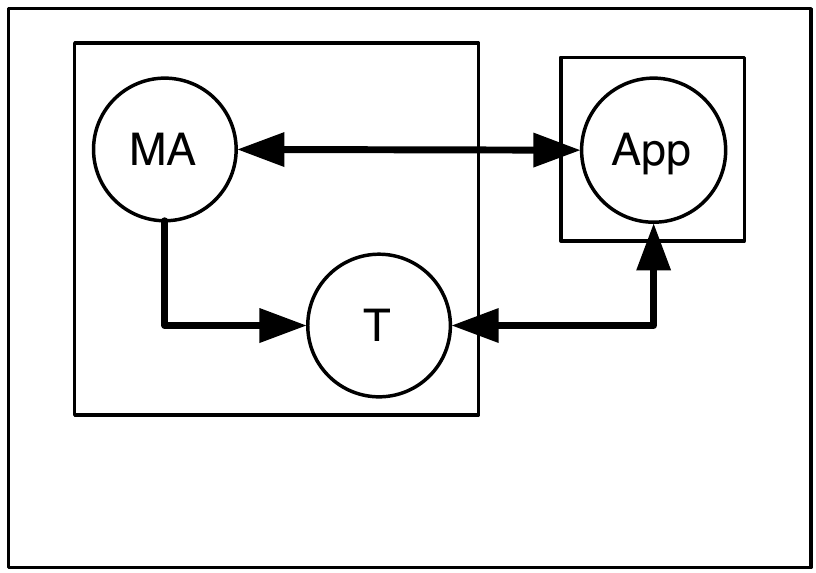}
          \caption{}
     \end{subfigure}
     \begin{subfigure}{0.22\textwidth}
          \includegraphics[width=\textwidth]{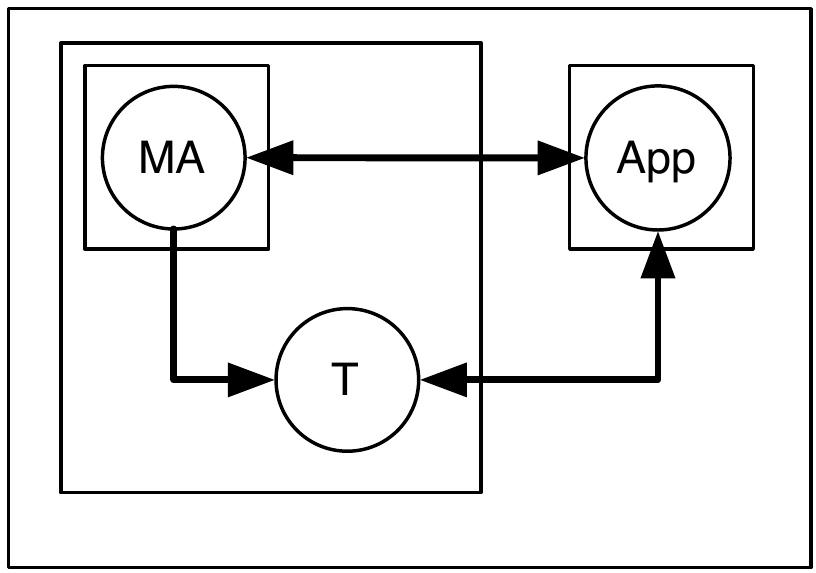}
          \caption{}
     \end{subfigure}
     \begin{subfigure}{0.22\textwidth}
          \includegraphics[width=\textwidth]{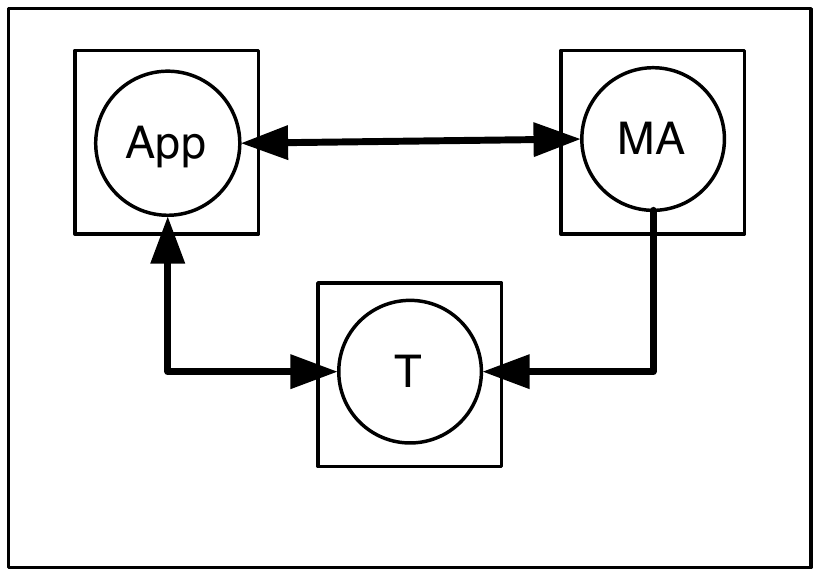}
          \caption{}
     \end{subfigure}
     \begin{subfigure}{0.22\textwidth}
          \includegraphics[width=\textwidth]{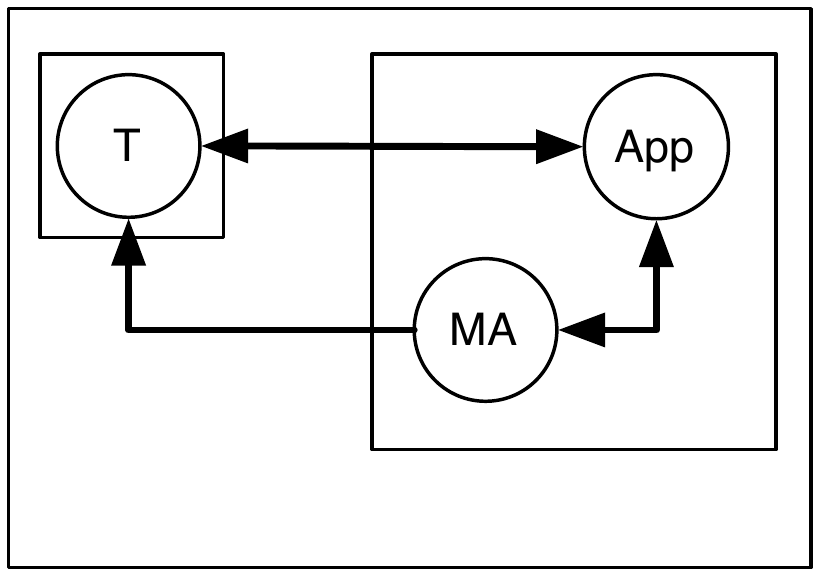}
          \caption{}
     \end{subfigure}
     \caption{Four possible layouts for an IMS \cite{lkim}. (a) The
       Measurement Agent (MA) and Target (T) share an execution
       environment; the Appraiser (App) is on a physically distinct
       host. (b) The Measurement Agent is isolated from the Target
       using dedicated hardware or virtualization; the Appraiser is on
       a physically distinct host. (c) Each component is on a distinct
       host.  (d) The Measurement Agent is hosted with the Appraiser
       while the Target is on a dedicated host. }
     \label{fig:ims-layouts}
     \end{center}
\end{figure*}

As shown in Figure \ref{fig:ims-layouts} (derived from \cite{lkim}),
these components may exist in a variety of 
architectures. Some systems may exist locally on a single platform,
while others may be components distributed across a network. Any
attempt at providing a general purpose interface for M\&A must enable
this kind of architectural diversity.

There are many types of evidence that can inform an integrity
evaluation.  One approach to integrity measurement, where measurements are 
cryptographic hashes of the static image of boot-time software,
is described by England et al. \cite{trusted-open-platform} and adopted by
the Trusted Computing Group (TCG) as part of the ``Trusted Boot''
technology \cite{tboot}. 
Other work, such as Sailer et al.'s Integrity
Measurement Architecture (IMA), takes cryptographic hashes of
executables and other files used
during platform operation \cite{ima}. Dynamic measurement
techniques, such as the Linux Kernel Integrity Measurer (LKIM), move
beyond hashes to introspection-based validation of critical runtime
data structures \cite{lkim}. 
Each of these measurement agents provides a unique form of evidence
that supports different notions of software integrity.

Similarly, there may be multiple attestation protocols which provide varying
degrees of protection and confidence in the evidence collected.
Many attestation protocols rely on a Trusted Platform Module (TPM)\cite{tpm} to
provide a static root of trust for measurement and the
measure-before-use sequence from the BIOS to the OS kernel and beyond
\cite{attestation-based-policy-enforcement}\cite{robust-integrity-reporting}\cite{secure-robust-integrity-reporting}. Flicker
defines a variation on this using the dynamic root of trust for
measurement to provide a protected, attestable execution environment
at runtime \cite{flicker}. OASIS proposes providing a protected
execution environment but roots trust in a physically unclonable
function (PUF) \cite{oasis}. Schemes for pure software-based
attestation typically rely on timing requirements for computing a
complex hash-like function over both the target and the measurement
agent itself \cite{armknecht}\cite{gardner}\cite{sake}\cite{pioneer}.


%% file: requirements.tex
\section{Properties of M\&A}
\label{sec:properties}

Loscocco et al. define a set of desirable properties for software integrity
measurement tools \cite{lkim}. Coker et al. define a similar set of desirable
properties for attestation protocols \cite{attestation-evidence-and-trust}. We
designed \sysname\ to support measurement and attestation components satisfying
these properties and, by extension, enable M\&A services to support trustworthy
inferences made by appraisers in a wide variety of possible trust decisions.

Two of the identified properties, \emph{completeness} and \emph{freshness},
relate specifically to measurement functions. Achieving these
helps ensure that appraisers are basing their decisions on measurement
data that reflects the current complete running state of the
target. One of the attestation properties, \emph{semantic
  explicitness}, addresses the degree that measurement data contains
the necessary evidence required for the trust decision that triggered
the attestation. These three properties 
govern the quality of trust decisions that the M\&A mechanisms
support. Therefore, it is imperative that an M\&A framework be
capable of supporting measurement agents and attestation protocols
with these properties for all anticipated trust decisions requiring
measurement evidence as an input.

In order to address the important question of whether the appraiser
should trust the evidence that it has received, three additional properties
apply: \emph{authenticity}, \emph{correctness}, and
\emph{protection}. Authenticity guarantees that the evidence came from an
authentic source and reflects the target of interest. Correctness relates to
the sound design and correct implementation of all of the M\&A component
mechanisms. Protection addresses the ability of the system to sufficiently
protect all of the M\&A components to enable them to meet their trust
obligations to the appraiser. These properties apply both to individual
measurement agents and attestation protocols, and to the M\&A framework itself. The
framework must be designed to (1) guarantee that the correct mechanisms are
invoked when necessary, (2) ensure order of operation where out of order
execution could impact trust, (3) limit accesses to and by M\&A components, and
(4) restrict access to M\&A functions and resources from other parts of the
system.

The purpose of an M\&A framework is to provide a general-purpose service for
integrity measurement. The properties of \emph{flexibility} and \emph{usability}
apply to this goal. Flexibility
implies an ability to (1) support multiple appraisers, measurement agents and
attestation protocols; (2) select appropriate mechanisms in the context of
given trust decisions, perhaps filtering results according to specific system
policies (e.g., privacy); and (3) incorporate future M\&A mechanisms. 
Usability implies that 
context should tailor the presentation of measurement data,
including possible post-collection processing, to the specific needs of a given
trust decision. 

\subsection{Critical Features}
\label{sec:properties:features}

We designed \sysname\ to support the eight high-level properties described
above: completeness, freshness, semantic explicitness, authenticity,
correctness, protection, flexibility, and usability. To meet these high-level
properties, \sysname\ includes the following concrete architectural features:

\begin{itemize}

     \item{\textbf{Support for Multiple Appraisers}: Platform integrity
       evaluations may be required in a wide variety of circumstances, and may
       be evaluated by several different appraisers. Each appraiser may
       have its own notions of what evidence is required to show integrity, and
       how that evidence should be collected and presented. A \emph{flexible}
       M\&A system should be \emph{usable} across a range of appraisers to
       prevent duplication and fragmentation of M\&A functionality.}
 
     \item{\textbf{Support for Multiple Attestation Protocols}: 
       Different attestation protocols can all present the same primary 
       evidence,
       but the appraiser may be able to draw different conclusions on the
       \emph{correctness}, \emph{freshness} and \emph{authenticity} of the data,
       based on the different bundling style. 
       For example, one protocol may produce an aggregate
       measurement report with a single signature covering all measurements,
       another protocol may produce individually signed reports for each piece of evidence
       collected, and a third may produce a chain of reports where each
       signature covers a report and the hash of the previous signed
       report. Allowing the policy-driven
       selection of protocols to suit the needs of each specific attestation
       scenario enhances the \emph{flexibility} and \emph{usability} of the
       M\&A system.}
        
     \item{\textbf{Support for Multiple Measurement Agents}: 
       Different evidence may
       be required to show the integrity of different components or to provide
       varying levels of confidence in one component. Some measurement agents
       may provide only boot-time evidence of integrity, such as a TPM quote,
       while others may generate \emph{fresher} or more \emph{complete}
       evidence showing integrity at runtime. Additionally, appraisers may
       require the use of a specific measurement agent to ensure
       \emph{correctness}, e.g., an appraiser may accept the reports of a
       malware scanner from one vendor but not another. It is important for the
       M\&A service to allow for policy-driven selection of measurement agents
       appropriate to a scenario. Like support for multiple appraisers and
       protocols, this feature directly supports the \emph{flexibility} and
       \emph{usability} of the M\&A system.}
       
     \item{\textbf{Policy-based Negotiation}: The M\&A service should be
       capable of negotiating with a peer to select a protocol and evidence set
       that satisfies each peer's policy. Appraisers are
       likely to seek the most detailed evidence of attester integrity
       possible, while attesters are likely to want to limit disclosure of 
       sensitive
       platform information. Both parties may want to limit the resources 
       required to complete the attestation. Support for policy-based 
       negotiation allows parties to find a protocol and evidence pair that is 
       consistent with these goals. The need for policy-based negotiation is 
       largely a consequence of support for multiple appraisers, protocols, and 
       agents and thus contributes to the \emph{flexibility} and 
       \emph{usability} of the system.}

     \item{\textbf{Discrete M\&A Functions}: To ensure the trustworthiness of
       their measurements, the service should ensure measurement agents
       are \emph{protected} from both the target of their measurement and from
       other parts of the measurement framework. For example, measurement
       agents may require exceptional authority, such as the ability to read
       kernel memory or attach a debugger to arbitrary processes. These
       privileges should be isolated in the smallest possible components and
       carefully controlled via platform access control policy. This isolation
       also allows for fine-grained
       policy regarding the \emph{authenticity} and \emph{correctness}
       properties of each component.}

     \item{\textbf{Support for Registration}: To provide extensibility of the
       supported attestation protocols and measurement agents, the service
       must include a registration mechanism to ensure that only valid
       combinations of agents and protocols are available for negotiation, 
       and that the platform's
       security policy is enforced for all installed components. Registration
       ensures that (1) only \emph{correct} components are accessible by the M\&A
       system, that (2) the purpose of these components is \emph{explicitly}
       defined, that (3) these components will be correctly \emph{protected},
       and that (4) their reports are \emph{authentic} because they can not be
       circumvented.}

     \item{\textbf{Portability}: The M\&A service should support deployment in a
       wide range of systems, including different components within a single
       system. In order to provide a \emph{complete} and
       \emph{protected} platform view, the M\&A service should be designed and
       implemented in such a way that it can function in any
       environment, including on a client system, server, or embedded device
       in physical systems; or in the host, hypervisor, guest, and/or dedicated
       virtual machines (VMs) in a virtualized system.}

     \item{\textbf{Composability}: Integrity evaluations may require the
       gathering and evaluation of evidence from multiple administrative realms
       within a single platform, or across multiple
       hosts. Multiple instances of the M\&A service should work
       together to delegate both evidence collection and evaluation tasks to
       the most appropriate instance. 
       Because measurement
       functionality must be \emph{protected} from the target of measurement,
       providing a \emph{complete} evaluation of a virtualized platform may
       involve cooperation between instances of the M\&A service in a guest VM,
       in an administrative VM, and in the virtual machine monitor (VMM) 
       itself.}

     \item{\textbf{Support for Complex, State-based Measurements}: The service
       should allow for complex evidence collection. An attestation may require
       invoking a collection of measurement agents that each inspects a
       different aspect of target state. To perform a \emph{complete}
       evaluation, collected evidence or intermediate evaluation results may
       introduce additional measurement requirements. For example, if a
       software inventory measurement shows that a Kerberos authentication
       package is installed, a measurement of the Kerberos configuration may be
       required. 
       These kinds of recursive measurement
       dependencies necessitate M\&A support for incremental discovery of
       measurement requirements.}
\end{itemize}

\begin{table}
\tiny
\begin{center}
\begin{tabular}{r||c|c|c|c|c|c|c|c|}
& \rothead{Completeness} & \rothead{Freshness} & 
\rothead{Semantic Explicitness} & \rothead{Authenticity} &
\rothead{Correctness} & \rothead{Protection} & \rothead{Flexibility} &
\rothead{Usability} \\
\hline
\hline
Multiple Appraisers  & & & &  &  & & X & X \\
Multiple Protocols   & & X & X& X & X & & X & X \\
Multiple Agents       & X & X & X& X & X & & X & X \\
Negotiation             & &   & & & & & X & X \\
Discrete Functions  & &   & & X & X & X &  &  \\
Registration             & &  & X & X & X & X &  &  \\
Portability                     & X & & & & & X & &  \\
Composability               & X & & & & & X & & \\
Complex Measurement & X & & & & &  & &  \\
\hline
\end{tabular}
\end{center}
\caption{Summary table showing how our critical features for an M\&A framework 
  align with the properties of M\&A identified by Loscocco et al. in
  \cite{lkim} and Coker et al. in \cite{attestation-evidence-and-trust}.}
\label{table:properties}
\end{table}

Table \ref{table:properties} gives a summary of how these features collectively
support all the identified properties for M\&A. 

\subsection{Adversary Model}

The goal of measurement and attestation is to detect adversaries that seek to
alter the long-term behavior of a platform by modifying long-lived data
residing either in memory or on persistent storage. Specific measurement
configurations may address adversaries with limited capabilities, such as
modification of user level programs but not kernel data. Other configurations
may target much more powerful adversaries capable of arbitrarily modifying data
anywhere on the platform, including in the OS, Virtual Machine Manager (VMM),
and BIOS/System Manage Mode (SMM) memory and storage. A framework that attempts
to unify all M\&A activities should support measurements that can be used in
aggregate to identify traces of all classes of adversaries.

The details of the measurement strategies employed by measurement agents are out
of scope for the purposes of this paper and the development of \sysname.
The properties and requirements described in this section allow for 
measurement agents to be used in conjunction to detect existing and future 
adversaries.

\sysname's emphasis on composability is particularly important to supporting
detection of a broad range of adversaries. \sysname\ is designed to support
collection of evidence at different levels of trust throughout the platform.
By evaluating evidence collected at a level of trust higher than the expected
adversary's capabilities, the appraiser can justify confidence in less trusted
components.


%% file: related-work.tex
\section{Related Work}
\label{sec:related-work}
A primary purpose of operating systems is to provide a set of reusable
abstractions to support the development and execution of user
applications. Features that are common across many
applications are good candidates for extraction into a system-wide
service to free developers from the need to maintain separate
solutions and to centralize administration for end users.

The Pluggable Authentication Modules (PAM) framework is a good example
of this philosophy \cite{pam}.  Authentication was originally
performed directly in the login process, but as the number of programs
implementing authentication and mechanisms for providing
authentication grew, the PAM library became necessary to provide a
single interface for programmers and a configuration point for users.
Thanks to this centralization, 
the system administrator uses a single configuration system to define 
what kinds of authentication may be used in what circumstances.

The need for a system-wide service for integrity evaluations follows a 
similar argument. There is a growing body of work on software integrity 
measurement and attestation.  We have observed numerous
instances of isolated or implicit integrity evaluations in common
platform usage, and a growth in the number of mechanisms supporting the 
collection and presentation of platform integrity. 

\subsection{Measurement Agents}
\label{sec:related:agents}
Measurement agents gather evidence that must be evaluated against some
policy to determine a target's integrity. 
There are a large number of measurement agents that collect evidence
that may be part of an integrity decision. TPM quotes are a common
form of evidence used to verify that the platform software was valid
at platform startup.  Other tools, such as IMA \cite{ima} and
Bit9 \cite{bit9}, provide load-time checks on programs as they launch
and are used to verify that the software being run was valid at
program start time. Dynamic runtime systems, like LKIM \cite{lkim} and
Semantic Integrity \cite{semantic-integrity}, can verify that both the
static and dynamic state of a piece of software is valid at a specific
time in the process's life cycle. Common system administrative tools
can also be used as measurement agents. For example, the firewall
configuration could be measured against an approved configuration, or
a recently completed virus scan could be used as evidence to inform an
integrity decision.

Some agents, such as LKIM, IMA, and tools for retrieving TPM quotes, may 
be designed to provide only the basic collection capability. Others, 
such as SecureBoot \cite{secure-boot} and IMA/EVM \cite{evm}, combine 
evaluation with collection. \sysname\ can incorporate the results of all
types of measurement agents, and provides a centralized framework for 
the selection and aggregation of measurements.

\subsection{Existing M\&A Frameworks}
\label{sec:related:frameworks}

Several M\&A frameworks, including Trusted Network Connect (TNC),
SAMSON, and OpenAttestation are implemented and in active use.  These
frameworks generally include their own measurement agents, a specific
protocol for gathering and communicating measurements, and components
for evaluating the evidence presented. These systems are somewhat
comparable to \sysname, but have much narrower focus and
limited extensibility. TNC is focused specifically on attestation for
the purpose of access control \cite{tnc}. SAMSON is focused on remote
attestation of client machines in an enterprise
environment \cite{samson}. And, OpenAttestation is focused on remote
attestation in enterprise or cloud environments \cite{openattestation}.
Table \ref{table:existing} shows how existing
framework solutions for integrity compare to \sysname\ in satisfying the 
features described in Section \ref{sec:properties:features}.  

None of these systems addresses how attestation can
be more fully integrated into the platform to reduce redundancy and
ease administration. As a result, on a typical GNU/Linux platform, two 
completely independent implementations of TNC may be installed: one as part
of the WPA\_supplicant tool for wireless network connectivity, and
another as part of the strongSwan IPsec package. These implementations
are derived from independent codebases, managed via distinct
configuration files, and utilize redundant but incompatible plugins
for integrity measurement collection and verification.

\begin{table}
\tiny
     \begin{center}
          \begin{tabular}{| l | *{7}{|c} |}
            \hline
            Feature                              & SAMSON & TNC & O.A. & \sysname \\
            \hline
            Multiple Appraisers           &      &  X &        & X \\
            Multiple Protocols             &      &     &        & X  \\
            Multiple Agents                 & X   &  X &        & X  \\
            Policy-Based Negotiation  &      &     &        & X  \\
            Discrete Components        &      &     &        & X  \\
            Registration                       &      &  /   &   & /  \\
            Portability                          &      &     &        & X  \\
            Composability           &      &   &         & X \\
            Complex Measurements    & X    & X &         & X  \\
            \hline
          \end{tabular}
     \end{center}
  \caption{Comparison of framework architectures for integrity measurement
                against the list of critical features delineated in Section 
                \ref{sec:properties:features}. `X' indicates a feature is 
                supported, `/` indicates a features is partially supported, 
                and a blank indicates no support for the feature.
                \label{table:existing}}
\end{table}

Nearly all systems discussed here provide a plugin mechanism to run
multiple types of measurement agents. The TNC specifications are
specifically designed around the idea of pluggable integrity
measurement collectors (IMCs) and integrity measurement verifiers
(IMVs). During attestation, the TNC client (attester) executes a
predefined set of IMCs, and the TNC server (appraiser) executes a
corresponding set of IMVs. The IMCs and IMVs communicate with
each other via the TNCCS protocol\cite{tnccs}.  SAMSON is designed
around a plugin architecture, and uses TNC interfaces and
protocols. OpenAttestation is the only system to hard-code the
supported measurement agents.

Few systems feature policy-based negotiation, and in most cases the
evaluator demands a particular set of measurements and the client must
either allow those measurements or refuse to participate. In
some commercial systems targeting enterprise client management, the
attesting software seems to be hard coded to unconditionally satisfy
any requests for attestation. No system we evaluated
supported selection of scenario-specific attestation protocols, or
collection and composition of measurements from multiple
administrative domains.  

Support for strong isolation is also remarkably rare. In all the
systems we examined, measurement collectors and evaluators ran
within the same process that responds to requests and
communicates with the attestation peer. 
Given that few systems support the other features, it is unsurprising
that they do not provide much support for registration. OpenAttestation 
does not support plugins, so there is nothing to register. TNC specifies
that each IMC/IMV filename be explicitly listed in a configuration
file, but this simple mechanism contains insufficient metadata about
each component to fully inform the types of decisions a framework must
make. The SAMSON documentation does not indicate any specific
registration mechanism. 

\subsection{Auditing Frameworks}
Network monitoring systems, such as Nagios \cite{nagios}, are intended
to allow system and network administrators to quickly collect status
data on a large number of systems. Like \sysname, auditing systems
tend to be built around a modular design, and may have modules for
collecting similar data such as system logs, TPM boot-time
measurements, or filesystem hashes. However, auditing systems are
generally aimed at collecting statistics
within a single administrative domain, and are designed around a
much simpler trust model. Auditing agents are installed and configured
by an administrator, and are specifically designed to report any
requested data to the single central auditor controlled by the network
administrator.

The need for auditing capabilities for cloud computing platforms is
widely recognized. Ko et al. \cite{trustcloud} identify security,
privacy, accountability, and auditability as components of trust in
cloud computing, and describe a high-level layered model, called
TrustCloud, for analyzing accountability in a cloud
environment. Abbadi et al. \cite{abbadi2012} provide a more detailed
breakdown of requirements for supporting trust in clouds, and define
TPM-based attestation protocols for one-way and mutual
authentication. They do not delve into how the required agents should
be implemented on the hosts or integrated into existing platform
services other than to note that the agents must not ``reveal domain
credentials in the clear, \dots\ transfer domain protection keys to
others, [or] \dots\ transfer sensitive domain content unprotected to
others''.  \sysname\ gives a concrete framework for implementing and
integrating these concepts.

Flogger \cite{flogger} implements an auditing system using a kernel
module for interception of filesystem accesses on a physical host and
within a VM, a process for uploading logs from VMs to a receiver
process on the physical host, and a database that consolidates reports
from multiple physical machines. This work is narrowly focused on the
problem of cloud auditing of file system events. It does not consider
other applications of attestation, and leaves security and integrity
concerns as an area of future work. Progger \cite{progger} is follow-on
work that expands the scope of Flogger and addresses some of the log
integrity concerns, but is still narrowly focused on data
provenance auditing, and relies on the integrity of VM kernels
to provide trust. Both Flogger and Progger are examples of systems that
assume a particular goal for evidence collection and implement a
single collection and presentation strategy to support that
goal. \sysname\ provides a design methodology for isolating these
concerns and a general infrastructure for integrating them into
systems.

Singh et. al \cite{iotmiddleware} identify the need for strong, 
policy-driven middleware for both audit collection and enforcement for
internet of things (IoT) and cloud-based systems.  While they focus on 
system-wide information
flow control, they call out specific need for trust on the platform and
ways to verify that auditing policy is being faithfully enforced.  
We see \sysname\ as complementary; it provides both the 
on-platform capability for policy enforcement and verification, and
a trusted remote attestation mechanism. 


%% file: architecture.tex
\section{Architecture}
\label{sec:architecture}

\begin{figure}
\begin{center}
\includegraphics[width=0.4\textwidth]{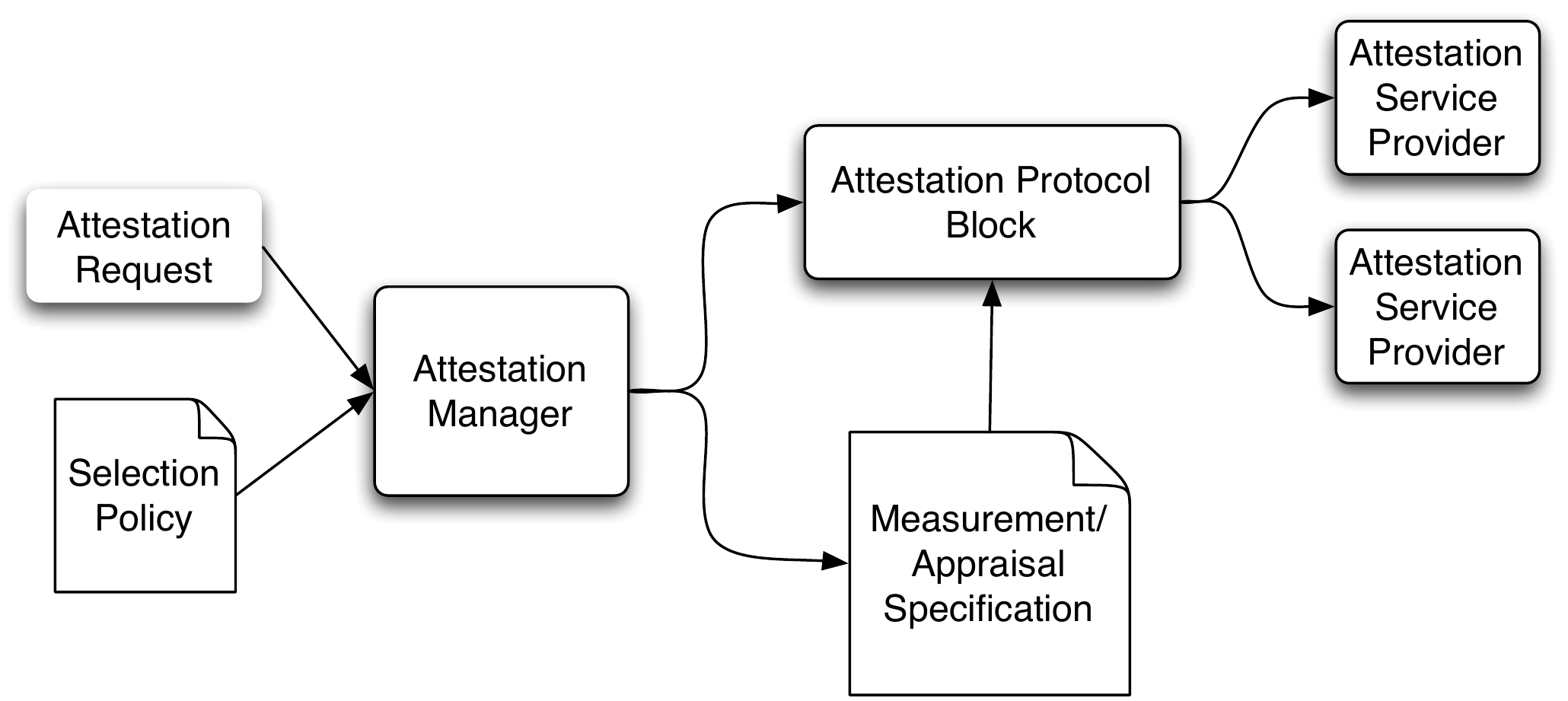}
\caption{The architecture of \sysname\ showing the Attestation Manager (AM),
  Attestation Protocol Blocks (APBs), and Attestation Service Providers (ASPs).
\label{fig:basic-arch}}
\end{center}
\end{figure}

\sysname\ is our prototype system that is explicitly designed to meet all of 
the desirable properties discussed in Section \ref{sec:properties}. 
Figure \ref{fig:basic-arch} shows the basic architecture of
\sysname. The \emph{Attestation Manager (AM)} (Section \ref{sec:am})
receives incoming requests, uses the \emph{Selection Policy} (Section
\ref{sec:selection}) to negotiate which protocols to run and what evidence to
gather, and spawns the agreed-upon \emph{Attestation Protocol Block (APB)}
(Section \ref{sec:apb}). Following the prescribed \emph{Measurement
  Specification} (Section \ref{sec:measurement-specification}), APBs invoke
\emph{Attestation Service Providers (ASPs)} (Section \ref{sec:asp}) to
gather required measurements.

\subsection{Attestation Roles}
Measurement, attestation, and evaluation are most commonly conceived as one
party, the \emph{attester}, generating evidence and presenting it to a remote
party, the \emph{appraiser}, for evaluation. Figure 3 depicts how this common
model of attestation is supported by \sysname. Prior to the flows
pictured, the appraiser AM receives a request to perform an integrity evaluation
of the attester. The attestation begins with a negotiation (flow 1)
between the attester and appraiser AMs. Once a suitable protocol/evidence pair
has been agreed upon, the AMs each execute the APB implementing their half of
the protocol (flow 2). During the protocol execution, measurements are
gathered on the attester by ASPs and passed to the APB (flow 3). The attester's
APB bundles the evidence and sends it to the appraiser's APB (flow 4). The
appraiser's APB parses the evidence received from the attester and passes it to
a series of ASPs representing the appraisal decision logic (flow 5). The result
of this appraisal is then returned, possibly with additional supporting data, to
the original requester.

\begin{figure}
\begin{center}
\includegraphics[width=0.4\textwidth]{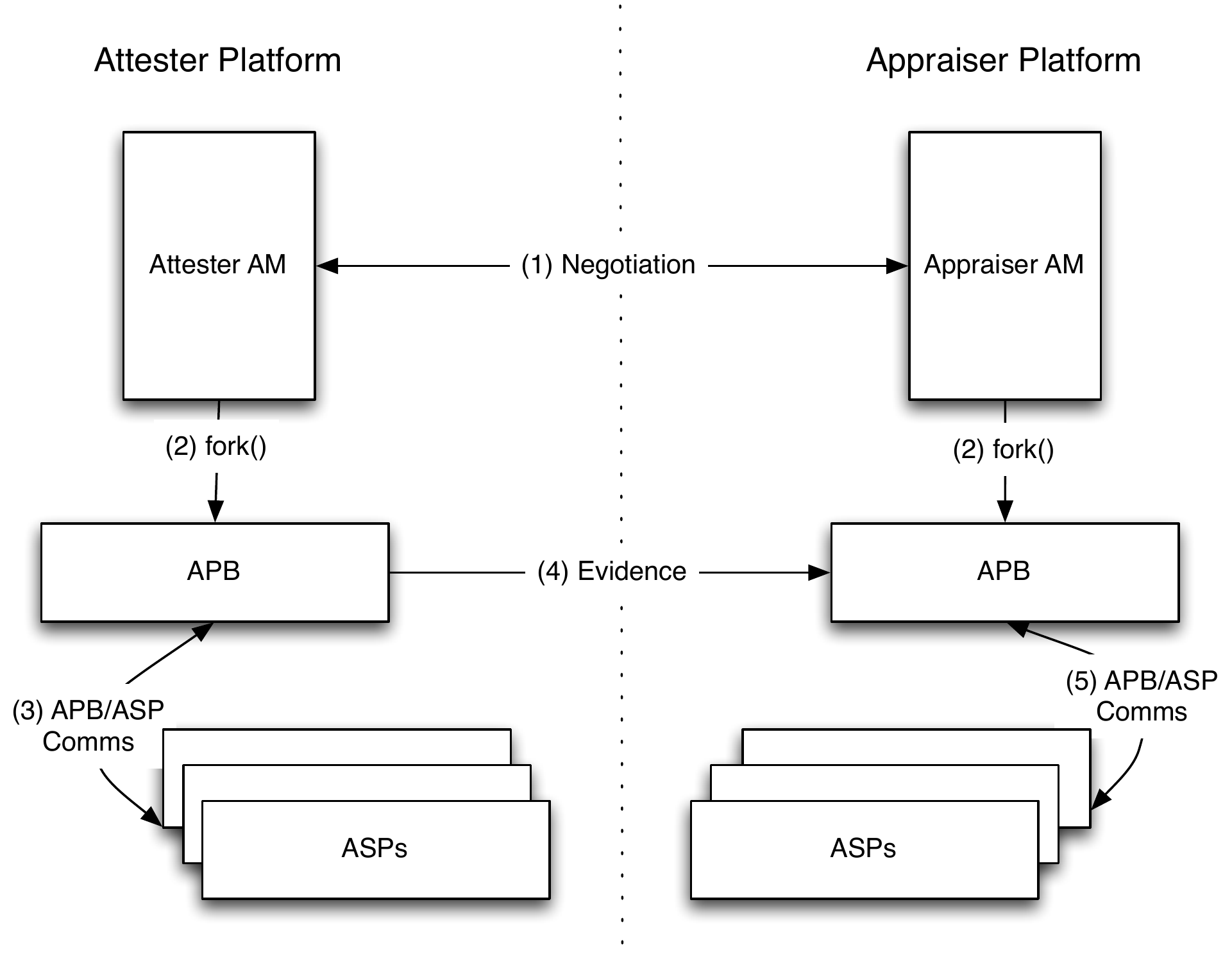}
\caption{A common model for attestations in which separate instances of 
  \sysname\ play the roles of attester and appraiser}
\label{fig:attester-appraiser}
\end{center}
\end{figure}

The diversity of scenarios
for integrity evaluation implies additional models for attestation. For example,
some may happen entirely locally to a platform; others may split the
appraisal, performing some local checks and some remote checks; and some
attestations may include third party trusted appraisers. \sysname\ is 
flexible enough to support these diverse models. In particular, any
component in \sysname\ may perform as either an attester or an appraiser, even 
within the same attestation.

\subsection{Security Model}
\label{sec:security-model}
Section \ref{sec:properties} discussed the importance of protecting an
attestation service from interference by the target of measurement. Attestation
services regularly communicate with untrusted parties, and require significant
local authority in order to collect measurements. This makes any attestation
service an attractive attack vector for adversaries seeking to gain access to a
platform.  To limit the impact of an adversary subverting an attestation
component, it is imperative that the entire system be sandboxed to the greatest
possible extent, and that internal components are protected from one another.

\sysname's security model is designed around discrete M\&A
functions. The AM, APBs, and ASPs each execute in separate processes to allow
OS or hypervisor level controls to assign only the necessary privileges to each
component. \sysname\ uses standard UNIX access control mechanisms, POSIX
capabilities, SELinux, and VM isolation and introspection to provide varying
levels of isolation between the components.

This isolation allows low-level measurements to be collected by more trusted
components and used to support the trustworthiness of higher-level
measurements. For example, a measurement of a VM's kernel and \sysname\ instance
collected via VM introspection may give an appraiser confidence in the validity
of measurements collected by the \sysname\ instance running in the VM. The
ability to chain measurements to gain confidence in higher-level measurement
functionality is critical to justifying trust in the \sysname\
framework. Ultimately, the chain should be rooted in a highly trustworthy
component such as a small, verified hypervisor running an embedded
``mini-\sysname'' instance and statically measured into a TPM at boot time.

We have implemented necessary hooks and policies for governing \sysname\
interactions using standard POSIX discretionary access controls (DAC), Linux
capabilities, and SELinux mandatory access controls. These models can be
coordinated to provide granular control over process privileges.  The DAC model
allows each component in \sysname\ to run with individual user
permissions specified at registration. This allows the externally communicating
components (the AM and APBs) to run as unprivileged users while only the
measurement gathering ASPs are run with higher privileges.
A set of Linux capabilities \cite{linux-capabilities} can
be specified at registration to further limit the administrative actions an ASP 
running 
as the super user may perform.  SELinux provides an even greater level of
isolation, as each APB and ASP can be given a unique SELinux domain with exactly
the necessary privileges.  

SELinux also provides guarantees that measurements are collected by the correct
component invoked in the correct way. The AM's executable is given an SELinux
type that has sole transition access to the correct domain for the AM, files
containing keys used to identify the AM are given an SELinux type that can only
be read by this domain, and APB images are given types that can only be launched
by this domain. On APB launch, SELinux forces a transition into a domain with
access to exactly the set of credentials and ASPs appropriate for that
APB. Finally, ASPs are run in domains with the minimal privileges necessary to
execute their particular function.  Combining trust in \sysname\ and SELinux
with a carefully constructed protocol allows an appraiser to conclude that the
measurements presented by the attester were collected by the correct
components.

\sysname\ also uses SELinux's category mechanism to isolate concurrent
attestations similar to how virtual machines are isolated from one another
under sVirt\cite{svirt}. The AM is initially provided with a large set of
categories. Each attestation session is handled by spawning a child process of
the AM to perform negotiation and then execute an APB. The parent AM gives each
child a unique set of categories in which to execute. The APB can then
similarly apportion its categories to ASPs as they are executed. This policy
protects the platform from subversion of \sysname, protects \sysname\ from
subversion of the platform (excepting attacks that subvert the operating system
kernel), and protects each component of \sysname\ from subversions in other
components.

For many measurement goals, OS level isolation is insufficient to guarantee the
needed isolation of measurement agent and target. Most notably, measurements of
an OS kernel itself can't be reliably performed from a process running on top
of the OS. One solution to this problem is to implement ASPs to self-protect
using existing hardware mechanisms. Specifically, an ASP could use a 
Flicker-like approach to establish a protected execution environment or Intel's Secure
Guard eXtensions (SGX) to create a secure enclave\cite{sgx}. These solutions
work and are supported by the \sysname\ architecture, but may require
substantial code replication across multiple ASPs.

\subsection{Multi-Realm Attestations}
\label{sec:multi-realm}

\sysname\ is intended to be replicated in each administrative or protection
domain throughout a platform. During negotiation and attestation, instances of
the architecture in one domain may delegate decisions or evidence collection to
another instance in a more appropriate domain. This scheme reuses the existing
\sysname\ functionality to provide a common interface for invoking measurement
capabilities that require greater isolation than can be provided within a single
operating system.

For example, on a virtualized platform there may be instances of the
M\&A architecture running:
\begin{itemize}
\item In each guest VM, for measuring the userspace of that VM
\item In the administrator VM, for measuring the kernels of the
guests and the administrator VM's userspace
\item In the VMM, for measuring the administrator VM
\end{itemize}
This hierarchy allows for trustworthy collection of evidence at all
levels from the guest VM to the kernel of the administrator
VM. Given appropriate hardware protection capabilities, another
instance capable of measuring the VMM itself is possible.

There are many open research challenges related to combining measurements from
multiple administrative domains that must be answered. These include: (1) the 
correct place to store policy and perform negotiations, (2) in what order to 
invoke measurements, and (3) how to endorse and combine measurements to produce 
a thorough argument that each measurement was properly collected and 
communicated.  \sysname\ is able to support many alternate
solutions to each of these problems, and thus provides a useful basis for
experimentation and the eventual integration of adopted solutions.

\subsection{Attestation Manager}
\label{sec:am}
The AM has two jobs in \sysname: (1) it acts as a registration point for APBs,
ASPs, Measurement Specifications, and (2) it is responsible for
negotiating and dispatching protocol/evidence pairs for each attestation
request.
The same AM
software is capable of acting as either an appraiser or an attester, and in
complex attestation scenarios may take on elements of each role.

In order to negotiate an attestation scenario with a peer in good faith, the AM
must know which APBs, ASPs, and Measurement Specifications are available on the
system.  We have partially implemented a registration mechanism for \sysname\
that achieves this by relying on the target system's native package manager to
correctly resolve dependencies between components and assign appropriate
permissions (user, group, SELinux label) to the installed files. Each component
is identified by a UUID that is specified in a metadata file installed as part
of the component's package. These UUIDs are used at runtime to resolve
dependencies, and are the basis for the AM's negotiation protocol.
 
While this implementation meets many of our goals such as dependency
tracking and automatic security label assignment, it does not allow
for additional checks on the pedigree of individual components nor does it
facilitate smooth updating of the AM's selection policy. Currently,
these checks are performed at load time using the metadata file provided with
each component. Future versions of \sysname\ will feature a more complete 
registration mechanism that performs these checks at registration time.
This mechanism will enhance the existing package management functions with
customized dependency and pedigree checks using the included metadata file.
Additionally, registration should move away from explicit UUID-based
dependencies in favor of feature-based dependency tracking. For example, APBs
should be able to specify measurement collection features by name and pedigree
requirements rather than specific UUIDs of required ASPs. This will allow for
negotiations based on properties of attestations rather than well known UUIDs.

\subsubsection{Selection Policy}
\label{sec:selection}

Negotiation of an attestation protocol and measurement specification is guided
by local selection policies at both the attester and the appraiser.  The goal
of negotiation is to select a protocol/evidence pair that satisfies the
integrity checks required by the appraiser without violating the privacy or
exceeding the computational limits of the attester. The appraiser's selection
policy defines what protocols and evidence are necessary to assess the
integrity of a given attester requesting access to a specified resource. The
attester's selection policy defines which protocols and evidence it is willing
to provide to a given appraiser in order to gain access to a given resource.

\begin{figure}
\begin{center}
\begin{subfigure}{0.4\textwidth}
\includegraphics[width=\textwidth]{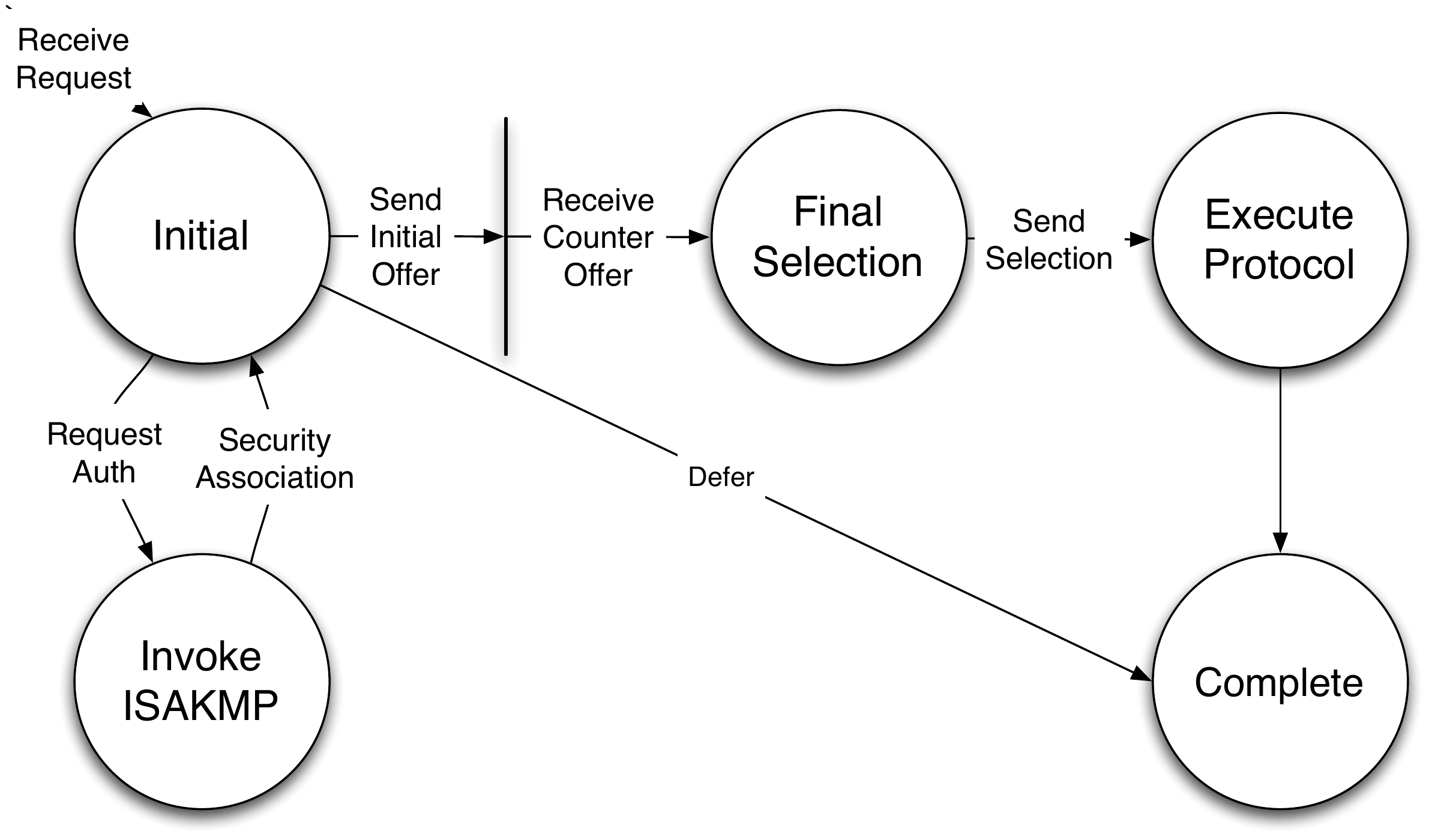}
\caption{Appraiser Selection Process}
\label{fig:selection-appraiser}
\end{subfigure}
\rule{0.4\textwidth}{0.4pt}
\vskip 6pt
\begin{subfigure}{0.4\textwidth}
\includegraphics[width=\textwidth]{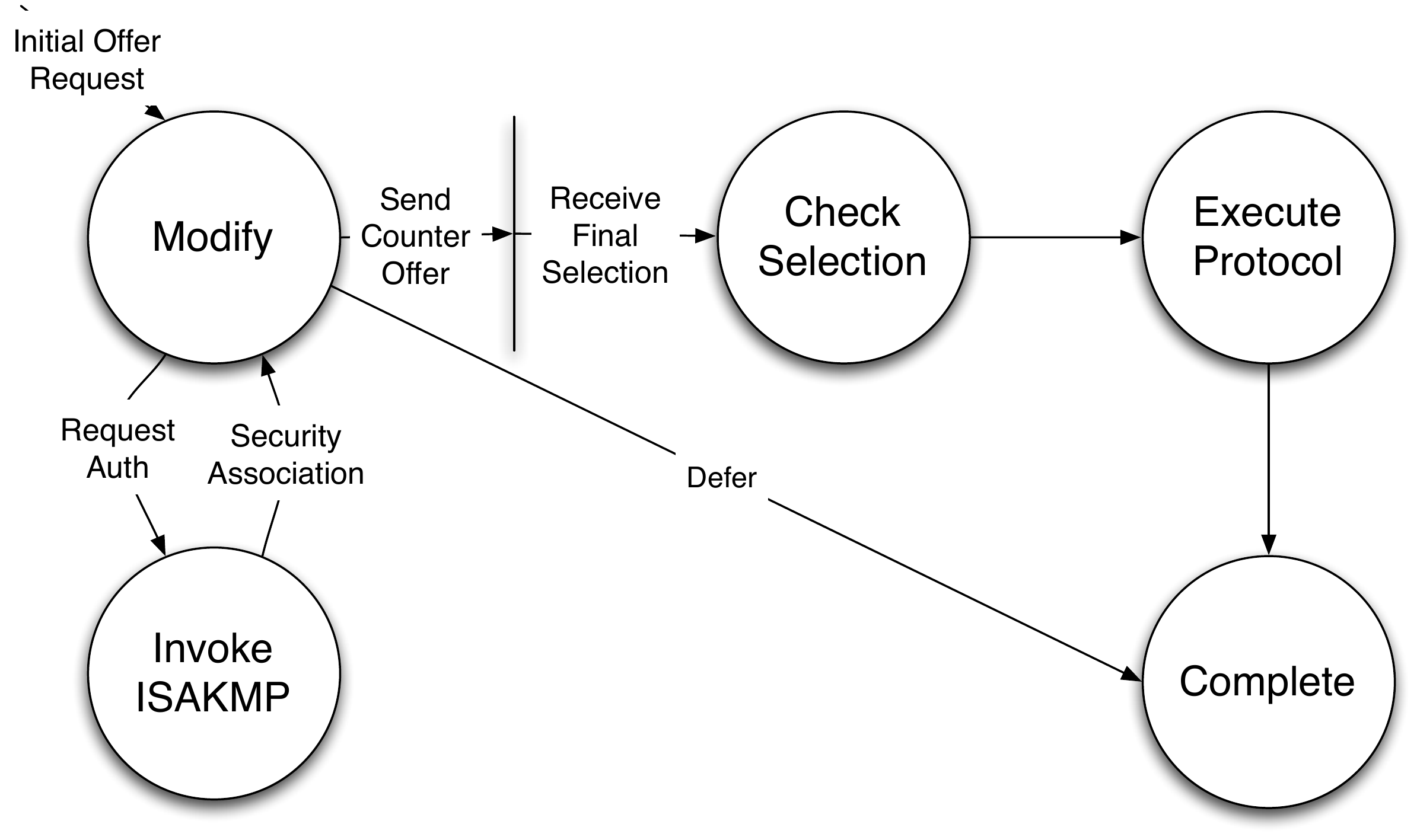}
\caption{Attester Selection Process}
\label{fig:selection-attester}
\end{subfigure}
\caption{State machine depictions of (a) Appraiser and (b) Attester selection
  processes. The selection policy is consulted at each state to determine what
  action to take.}
\label{fig:selection}
\end{center}
\end{figure}

The policy must contain a series of declarative rules mapping inputs describing
the current attestation to actions.  The policy may be stored either in a
simple file or as a database for larger policies.  For both the appraiser and
attester, the inputs to the selection policy are:
\begin{itemize}
\item the role of the party in the attestation
\item the other party's identity
\item the strength of this identity association
\item the resource being requested
\item the current state of the negotiation.
\end{itemize} 
Figures \ref{fig:selection-appraiser} and \ref{fig:selection-attester} show
state machine views of the negotiation and selection processes on the appraiser
and attester respectively.

A request for integrity evaluation received by the appraiser must specify the
identity of the attester and the resource being guarded.  The appraiser matches
these inputs against its policy, which may result in a match failure (not
pictured), a request for a stronger identity binding, a deferral, or a set of
protocol/evidence options.

In the first case, attestation is aborted and an error is returned. If a
stronger identity binding is needed, a call can be made to an ISAKMP daemon to
produce the needed association before continuing the negotiation. If
negotiation is deferred to another appraiser, the request is forwarded and a
proxy process may be created to forward messages if necessary. If a set of
options is returned, an initial contract is generated and sent to the attester.

The attester uses these inputs to consult its policy. If successful, this 
results in a counter offer: 
a subset of the offered options that the attester is willing to perform. 
These options are sent to the appraiser, which then 
consults its policy to determine the
preferred option, sends this option to the attester, and begins execution of
the corresponding appraisal APB. The attester performs a final policy check 
to verify the suitability of the appraiser's selection, then 
executes the selected APB.

This negotiation process is intended to ensure that the optimal protocol and
evidence are selected. Notably, the selection is made by the appraiser rather
than the attester. This necessitates an extra communication round trip in the
protocol, but ensures that the appraiser's prioritization is respected. An
alternate solution could treat the initial options as a prioritized list rather
than an unordered set, and trust that the attester will select the highest
priority option consistent with its policy. This choice does not seem to have
any direct impacts on the trustworthiness of the selection: the appraiser
shouldn't offer any options that are not sufficient for its trust goals, and it
must already trust that the attester is correctly implementing
negotiation. However, the extra step provides more explicit appraiser control
over the outcome of negotiation in exchange for minimal performance overhead
(performance is typically dominated by the measurement collection) that could
be easily eliminated by a cache of negotiation outcomes.

\subsection{Attestation Protocol Blocks}
\label{sec:apb}
APBs are responsible for understanding the requirements of a particular
attestation scenario as defined by the Measurement Specification, executing an
appropriate sequence of ASPs to satisfy the scenario, and collecting
the results generated by individual ASPs into a cohesive whole that is
consumable by the remote party. Dually, an APB may implement an appraisal
component that evaluates evidence by invoking a sequence of ASPs to verify
properties of the measurement data and synthesizing a final report indicating
the overall determination of integrity along with any required supporting
evidence. Protocols may be implemented either as two separate APBs, one for the
attester and one for the appraiser, or as a single APB that determines which
role to execute based on context provided by the AM. 

APBs rely on ASPs and/or other APBs to produce or evaluate measurements. Upon 
registration with the AM, each APB must provide an XML metadata file that 
statically lists supported Measurement Specifications and defines the set of ASPs
and sub-APBs required to execute the protocol. With this information,
the AM can ensure that all dependencies can be satisfied, and can invalidate a 
protocol if any of its dependencies are de-registered. 

We separate evidence collection and collation into ASP and APB functionality
respectively to allow for greater reuse of components, to support finer-grained
policy decisions, and to enable more granular access control decisions to
isolate pieces of M\&A functionality.  However, \sysname\ is flexible and can
accomodate ``fat'' APBs that collect evidence directly or ``fat'' ASPs that
collect multiple types of data. 

\subsection{Measurement Specifications}
\label{sec:measurement-specification}
Measurement Specifications define what evidence the requester requires
for a specific scenario.  Separating the evidence requirements from the
protocol used to collect and transmit evidence (APBs) allows the construction
of generic APBs that can be re-used for multiple attestation scenarios. Like
APBs, Measurement Specifications are registered with the AM and are identified
by a well-known UUID.  Once an APB/Measurement Specification pair is negotiated,
the APB is launched using the Measurement Specification as input.

Measurement Specifications contain as much information as is necessary to define
the type of evidence required. Implementations may define a specification 
language that provides rich syntax for defining complex evidence relationships.
Any such language must be understood by the APBs, which parse the
specification into a series of actionable instructions. \sysname\ includes
an implementation of a specification language as an optional
library. Specifications in this language define a set of measurement variables
that identify particular data requiring measurement, and measurement
instructions that define what measurements are required for variables of a given
type. During evaluation, measurement instructions may introduce new variables 
that must be measured. The complete measurement requires recursively evaluating 
these measurement obligations until no new variables are introduced.

\newcommand\mapbind[0]{>>=}

\begin{figure}
\footnotesize
\begin{tabular}{llll}
1 & \multicolumn{2}{l}{\texttt{measure p :: path}} &\\
2 &\hskip 2em &\texttt{| is\_reg p} &\texttt{= sha1sum p} \\
3 & & \texttt{| is\_dir p} &\texttt{= dirlist p \mapbind} \\
   & & &\hskip 1.3em \texttt{measure} \\
4 & & \texttt{otherwise} &\texttt{= success}\\
5 & \multicolumn{3}{l}{\texttt{do measure ("/etc" :: path)}}\\
\end{tabular}
\caption{Example measurement specification for recursively enumerating and
  hashing regular files in the \texttt{/etc} directory. 
}
\label{fig:spec-example}
\end{figure}

The complete syntax and semantics of \sysname's measurement specification
language are beyond the scope of this paper. The example 
given in Figure \ref{fig:spec-example} provides a
reasonable overview of the language's features. This example implements a
common goal of integrity measurement systems: to provide a
TripWire\cite{tripwire}-like summary of the hashes of all files recursively
found in the \texttt{/etc} directory. 

The specification language supports specification composition, since multiple
specifications may cause different evidence to be collected for the same piece
of target state. For example, another specification that extracts a list of users from
the \texttt{/etc/passwd} file may be combined with the example in Figure
\ref{fig:spec-example} to generate a measurement containing both a hash of
\texttt{/etc/passwd} and the list of users. The evaluation order of measurement
instructions is not strictly defined. Our example APBs utilize a queue of
measurement obligations and continue executing until the queue is empty, but
any strategy that guarantees that all obligations are eventually discharged is
valid.

\subsection{Attestation Service Providers}
\label{sec:asp}
Attestation Service Providers (ASPs) are the basic functional unit of
\sysname. Each ASP performs a specific, discrete function in evidence
collection tasks. For example, an ASP can gather a
specific piece of evidence from the system, ingest some type of evidence and 
contribute to an assessment of the target's integrity, provide post-processing 
functions such as
hashing or compression, or call out to external components such as another AM
or service. \sysname\ makes no distinction between ASPs that
collect evidence and ASPs that evaluate evidence. This decision is intended to 
help support more complex
attestation scenarios in which partial appraisals may occur locally with
measurement collection, or with a third party appraiser not involved in the
initial negotiations.

ASPs are invoked by an APB, using an implementation-defined interface.  In line
with our desire for discrete M\&A functions, ASPs 
run in their own address space and may have their own fine-grained policies. 
ASPs may be invoked as a discrete event, chained into a pipeline, 
or called multiple times 
for a single attestation. The ordering of invocation is determined by the APB 
and its interpretation of the 
Measurement Specification.



%% file: deployment.tex
\section{Example Use Cases}
\label{sec:deployment}
The prototype implementation of \sysname\ allows us to verify that
the proposed architecture is flexible enough to integrate and/or
consolidate integrity information in a variety of deployment
scenarios.  These include both augmenting existing systems with
additional plugins that call out to \sysname\ externally, and subsuming
existing services into \sysname\ as APBs or ASPs.  We provide
several examples of integrating \sysname\ in the following 
sections.

\subsection{Authentication}
\label{sec:scenario:auth}
Many UNIX systems use the PAM library to provide a system-wide
authentication service.  As discussed previously, PAM centralizes
authentication decisions, while providing an extensible architecture
through the use of plugin modules.  This extensibility makes
incorporating integrity information into authentication decisions
straightforward.

An adversary can easily use spoofed authentication windows to trick
users into giving up their credentials\cite{spoof-auth}. 
To enable users to confirm the integrity of a system before authentication,
we integrated \sysname\ with
authentication by creating a PAM module that calls out to \sysname\
for an integrity analysis. \sysname\ uses a corresponding PAM policy
that checks a system for compliance before the user enters his or
her credentials. If the system fails to meet the required policy, the
interface displays an alert, informing the user that the system is
noncompliant. The combination of PAM and \sysname\ made it trivial to
integrate system integrity verification into the login process.

\subsection{Network Access Control and IPsec}
\label{sec:scenario:nac}
Network access control (NAC) is implemented by the 802.1x standard,
which frequently uses the Trusted Network Connect (TNC) framework for
gathering measurements from the system before allowing access.  While
the limitations of TNC have already been discussed, the fact that TNC
is spiritually similar to \sysname\ allows interesting opportunities
for integration.

Our first method of integration was achieved by creating a TNC IMC/IMV
pair that calls out to the M\&A service, and uses the result of the M\&A
decision as input into TNC's overall NAC decision. This form
of integration is straightforward; however, it makes the M\&A service
a subordinate protocol to TNC, and limits the ability to use the richness
available when using the M\&A service directly. 

Our second integration of TNC incorporated the TNC Server,
TNC Client, TNC IMC/IMV interface, and the TNC communications protocol
into a pair of APBs, which are negotiated and directly launched by
\sysname's AM.  To achieve this, we took an
existing implementation of the above services (strongSwan) and wrote a
small shim layer to adapt the native interface of these services to
\sysname's APB interface. The resulting APB allows the TNC
infrastructure to be directly employed through the M\&A service for
any purpose, not just for NAC.

To trigger \sysname\ from the NAC process, we wrote a vendor-specific
Extensible Authentication Protocol (EAP) method for the NAC server and
client, hostapd and wpa\_supplicant respectively. The EAP method
communicates with \sysname\ via a UNIX domain socket, and serves as a
proxy between the attester and appraiser instances. Upon receiving a
request, the attester and appraiser undergo their standard negotiation
to select the appropriate attestation protocol and evidence for the
scenario. For TNC, we use a mapping of evidence spec UUID to sets of
IMC/IMV pairs to determine which IMCs/IMVs the APB should load and
execute. This allows the same TNC APB to provide and/or evaluate
different sets of evidence, depending on the result of negotiation.

\sysname's inherent flexibility allowed us to implement both methods 
of integration.  This flexibility is critical to supporting legacy systems
and provides multiple, straightforward migration paths for network
administrators to continue leveraging existing tools while taking advantage 
of \sysname's desirable properties.

TNC can also be used during IPsec authentication. Our integration with
TNC facilitated integration with IPsec. IPsec
negotiates security associations (SAs) between two hosts on an IP
network. These SAs allow for cryptographically authenticated and/or
encrypted traffic to be passed between the two hosts using information
incorporated into the SA.  To integrate \sysname\ with IPsec, we
modified the code for our vendor-specific EAP method to call out from
the strongSwan IPsec stack to the \sysname\ AM, which then chooses the
same TNC APB used for NAC to take a measurement and send results to
the IPsec service.

The ability to use the same TNC APB for both IPsec
and NAC shows that \sysname\ can consolidate measurement agents,
eliminating the need for custom TNC code in the NAC and IPsec
services. Further, the separation of protocol selection from required
evidence in \sysname\ allows the IPsec scenario to employ different IMC/IMVs 
than were selected for NAC.

\subsection{Host Monitoring}
\label{sec:scenario:monitoring}
Network administrators often want to make use of host integrity data as
part of a network monitoring system. \sysname\ includes a simple web
application that allows network administrators to manage host
identities, view historical integrity reports, and request fresh
integrity evaluations. Using the web interface, the
system administrator can register a host for monitoring 
and request fresh evaluations by specifying the host(s)
to be evaluated and monitoring criterion. The web application uses an
instance of \sysname\ to negotiate with 
each target, perform the evaluation, and generate a detailed report that 
is then stored in a database. 
This use case is common in an enterprise
deployment but is typically handled using ad-hoc or proprietary
reporting systems. \sysname\ provides the same benefits for a host
monitoring system that it does for access control focused use cases.

For broader impact assessment, an early version of \sysname\ is
currently deployed on a large scale enterprise network collecting,
archiving, and appraising both periodic and on-demand measurements
from several hundred actively used Linux-based systems.  It provides a
single, secure interface for collecting a wide variety of information,
including TPM quotes, software inventories, file integrity checks, and
kernel integrity measurements.  The IT system administrators expressed
appreciation for \sysname's ability to collect this variety of
measurements. They use five or more independent, proprietary systems
to collect the same information for other operating systems. Each of
these systems requires financial investment, training, configuration,
and introduces new security concerns to the IT architecture. Unifying
these features in \sysname\ eliminates this complexity and allows
administrators to quickly access the data they need in a consistent
interface.

\subsection{Internet of Things}
\label{sec:scenario:iot}
We further tested \sysname's flexibility and extensibility by
modifying it to measure non-traditional platforms, such as those used
as part of the internet of things (IoT).  Separate from the enterprise
use case, these platforms present significant challenges for M\&A such
as lack of common software bases, ad-hoc communications mechanisms,
and potentially severe resource constraints.  However, IoT devices
represent a growing threat~\cite{risky} to the security of the systems
with which they interact and thus need the same integrity evaluation
capability available to non-IoT systems. We divide the class of smart
devices into two categories: high- and low-capability.

High-capability devices are embedded systems that contain enough
resources and platform features to allow running the entire \sysname\
stack within the constraints of their environment.  Examples of such
systems include high-end smart appliances and gateways which run
stripped-down versions of standard multitasking operating systems such
as Windows or Linux.  These platforms have computing capacity, power,
storage, and memory protection necessary to run the entire \sysname\
stack, assuming that \sysname\ were ported to the target platform and
contained appropriate ASPs to take meaningful measurements.  We
demonstrated this by successfully running \sysname\ on a 32-bit ARM
Cortex-A series development platform and a commodity MIPS-based router
running OpenWRT~\cite{openwrt}.  Both platforms run on custom versions
of the Linux operating system.  Once configured, these high-capability
IoT devices operated as part of the \sysname\ system just as every
other platform did.

Low-capability devices are much more constrained, and often lack the
resources to run \sysname\  effectively without 
modification. As fully discussed by Clemens et al.~\cite{iota},
\sysname\ can be modified to provide negotiation and collation
features (essentially the AM and APB aspects) on a high-capability
device while deferring specific measurement collection to a
miniaturized instance running on the low-capability device. The
miniaturized instance lacks many of the protective features of
\sysname\, but implements the same interface for invoking measurement
capabilties. The end result shows how \sysname\ can be used to provide
centrally-managed, policy-driven, comprehensive, and efficient M\&A
capabilities to a broad range of platforms, from high-end enterprise
systems to severely resource-constrained environments.



%% file: future-work.tex
\section{Further Research Challenges}
\label{sec:future-work}

While \sysname\ is designed to satisfy the properties described in
Section~\ref{sec:properties}, there remain many open research
challenges.  Some of these challenges will be addressed by improving
the prototype implementation, others are broader challenges that require
further research.

As noted in Section~\ref{sec:am}, the existing registration mechanism is not 
expressive enough to support all of the desired protection properties, and
will need to be improved in future implementations.  We also believe that our 
current selection policy mechanism will need refinement to both the language
and the dispatch mechanism to allow for decisions based on attributes of an
attestation scenario that are not currently exposed.  
Improvements to registration and selection would also enable a richer
negotiation process based on attributes of the protocols and evidence
being selected rather than on UUIDs. Developing an attribute language
that is rich enough to encompass all possible measurements, protocols,
and trust properties while being precise enough to guarantee
compatibility between endpoints is a significant research problem.

As discussed in Section \ref{sec:multi-realm}, support for multi-realm
attestations is a major area of future work. \sysname\ is designed to
be replicated in each administrative domain. However, important
research questions remain that could impact the trustworthiness of
attestations. These include where to store policy and perform
negotiations, in what order to invoke measurements, and how to endorse
and combine measurements to produce a thorough argument that each
measurement was properly collected and communicated. These are all
open questions that must be answered for trustworthy multi-realm
attestations. \sysname\ provides a basis for experimenting with
alternate solutions to each of these questions.

Further development also suggests the need for a system to provide
comprehensive trust decisions as a service (TDaaS) in software systems. This
requires integrating \sysname\ into a larger framework that also
provides authentication, identity management, establishment of
security associations, and maintenance of a cache of known clients and
a history of previous successful negotiations.

As implemented, \sysname\ targets traditional computers running a
POSIX-compliant operating systems. As noted in Section
\ref{sec:scenario:iot}, \sysname\ has also been used to demonstrate
M\&A on sample IoT devices. However, the internet consists of many
different kinds of platforms, with varying hardware resources and
operating systems.  Supporting the separation guarantees and
negotiation requirements may be a straightforward engineering task for
many platforms, but others are so resource constrained that they
require further research and re-thinking of what is possible and
necessary for integrity measurement. Finally, computing platforms that
are not general purpose often contain custom software or firmware,
requiring that special, non-portable versions of \sysname\ be
integrated with the custom system.  Such diversity remains a challenge
to any framework that aims for broad adoption.


%% file: conclusion.tex
\section{Conclusion}
\label{sec:conclusion}

The time has come for a central framework for the collection and presentation of
integrity measurements for use in trust decisions. 
We believe that such a system must adhere to the properties
enumerated earlier in this paper, and prove that such a system is attainable
today with a discussion of \sysname, our prototype measurement and attestation
(M\&A) framework.

\sysname\ supports an array of attestation scenarios, measurement
types, and protocols, and it has a high degree of flexibility to
enable seamless integration with legacy systems. \sysname\ cooperates
with both the target and the requester in an attestation
scenario. Policy-based negotiation allows the requester to specify the
evidence required to complete an attestation, and it allows the target
to specify under what conditions each piece of evidence may be
released. We have demonstrated the applicability of \sysname\ through
multiple deployment scenarios, and integration of \sysname\ with PAM,
NAC, IPsec, IoT, and client monitoring.

It is our desire for \sysname\ to be used as a basis for further research
into the field of trustworthy integrity measurement as well as a starting
point for robust, system level M\&A services.
